\pdfoutput=1

\RequirePackage{fix-cm}

\documentclass[onecolumn]{svjour3}

\smartqed

\usepackage{graphicx}
\usepackage{amssymb,amsmath}
\usepackage{txfonts}
\usepackage[numbers]{natbib}
\usepackage{hyperref}

\hypersetup{pdftitle = The title of my PDF, pdfauthor = My name, pdfsubject= The subject, pdfkeywords = keyword1 keyword2 keyword3}
\hypersetup{colorlinks = true, linkcolor = blue, anchorcolor = red, citecolor = blue, filecolor = red, pagecolor = red, urlcolor = blue}

\journalname{Journal of Applied Mathematics and Computing}

\begin{document}

\title{Investigating the Newton-Raphson basins of attraction in the restricted three-body problem with modified Newtonian gravity}

\author{Euaggelos E. Zotos}

\institute{Department of Physics, School of Science, \\
Aristotle University of Thessaloniki, \\
GR-541 24, Thessaloniki, Greece\\
Corresponding author's email: {evzotos@physics.auth.gr}}

\date{Received: 26 August 2016 / Published online: 6 October 2016}

\titlerunning{Basins of attraction in the PCRTBP with modified Newtonian gravity}

\authorrunning{Euaggelos E. Zotos}

\maketitle

\begin{abstract}

The planar circular restricted three-body problem with modified Newtonian gravity is used in order to determine the Newton-Raphson basins of attraction associated with the equilibrium points. The evolution of the position of the five Lagrange points is monitored when the value of the power $p$ of the gravitational potential of the second primary varies in predefined intervals. The regions on the configuration $(x,y)$ plane occupied by the basins of attraction are revealed using the multivariate version of the Newton-Raphson iterative scheme. The correlations between the basins of convergence of the equilibrium points and the corresponding number of iterations needed for obtaining the desired accuracy are also illustrated. We conduct a thorough and systematic numerical investigation by demonstrating how the dynamical quantity $p$ influences the shape as well as the geometry of the basins of attractions. Our results strongly suggest that the power $p$ is indeed a very influential parameter in both cases of weaker or stronger Newtonian gravity.

\keywords{Three-body problem $\cdot$ Non-linear algebraic equations $\cdot$ Basins of attraction $\cdot$ Fractal basin boundaries}

\end{abstract}

\section{Introduction}
\label{intro}

The field of numerically solutions of algebraic systems of equations has a long history (e.g., \cite{A16,AA14,AAM16,EAA15}). At this point, we should emphasize, that all the above-mentioned references are exemplary rather than exhaustive taking into account the vast area of the field of applied mathematics.

In dynamical systems knowing the basins of attraction associated with the equilibrium points is very important since this knowledge reveals some of the most inartistic properties of the system. The sets of initial conditions on the configuration $(x,y)$ plane which lead to a specific equilibrium point (attractor) define the several attraction regions (known also as basins of convergence). Over the last years, the basins of attraction in several types of dynamical systems have been numerically investigated. In \cite{D10} the Newton-Raphson iterative method was used in order to explore the basins of attraction in the Hill's problem with oblateness and radiation pressure. In the same vein, the multivariate version of the same iterative scheme has been used to unveil the basins of convergence in the restricted three-body problem (e.g., \cite{Z16}), the electromagnetic Copenhagen problem (e.g., \cite{KGK12}), the photogravitational Copenhagen problem (e.g., \cite{K08}), the four-body problem (e.g., \cite{BP11,KK14}), the ring problem of $N + 1$ bodies (e.g., \cite{CK07,GKK09}), or even even the restricted 2+2 body problem (e.g., \cite{CK13}) have been studied.

The Newton-Raphson method, and of course the corresponding multivariate version of it, may be a very simple computational tool for numerically solving system of equations however it is not the most robust algorithm. Being more precisely, its main inefficiency lies to the fact that the convergence of the algorithm is greatly affected by the particular initial guess (e.g., \cite{FARA14,FAR15,JK98,SSY08}).

Undoubtedly, one of most important and intriguing topics in celestial mechanics as well as in dynamical astronomy is the classical problem of the planar circular restricted three-body problem (PCRTBP). This problem describes the motion of a test particle with an infinitesimal mass inside the gravitational field of two primary bodies which move in circular orbits around their common center of gravity \cite{S67}. The applications of this problem expand in many fields of research from chaos theory and molecular physics to planetary physics, stellar systems or even to galactic dynamics. This clearly justifies why this topic still remains very active and extremely stimulating.

In the present article we shall use the mathematical model of the planar circular restricted three-body problem with modified Newtonian gravity. In particular, the power $p$ of the gravitational potential of the second primary will vary in predefined intervals thus allowing us to examine two different scenarios: (i) the case of weaker Newtonian gravity, when $p < 1$ and (ii) the case of stronger Newtonian gravity, when $p > 1$. In both cases, our aim will be to determine how the shape and geometry in general of the Newton-Raphson basins of attraction are affected by the change in the value of the power $p$ of the potential.

The structure of the paper is as follows: In Section \ref{mod} we describe the basic properties of the considered mathematical model. In section \ref{lgevol} the evolution of the position of the equilibrium points is investigated as the value of power of the gravitational potential varies in predefined intervals. In the following Section, we conduct a thorough and systematic numerical exploration by revealing the Newton-Raphson basins of attraction and how they are affected by the value of the power $p$. Our paper ends with Section \ref{disc}, where the discussion and the conclusions of this work are presented.

\section{Presentation of the mathematical model}
\label{mod}

Let us briefly recall the basic properties of the PCRTBP \cite{S67}. The two main bodies, called primaries move on circular orbits around their common center of gravity. The third body (also known as test particle) moves in the same plane under the gravitational field of the two primaries. It is assumed that the motion of the two primaries is not perturbed by the third body since the mass of the third body is much smaller with respect to the masses of the two primaries.

The units of length, mass and time are taken so that the sum of the masses, the distance between the primaries and the angular velocity is unity, which sets the gravitational constant $G = 1$. A rotating rectangular system whose origin is the center of mass of the primaries and whose $Ox$-axis contains the primaries is used. The mass ratio is $\mu = m_2/(m_1 + m_2)$, where $m_1 = 1 - \mu$ and $m_2 = \mu$ are the dimensionless masses of the primaries with $m_1 > m_2$, such that $m_1 + m_2 = 1$. The centers $P_1$ and $P_2$ of the two primaries are located at $(-\mu, 0)$ and $(1-\mu,0)$, respectively.

The total time-independent effective potential function in the rotating frame of reference is
\begin{equation}
\Omega(x,y) = \frac{(1 - \mu)}{r_1} + \frac{\mu}{r_2^{p}} + \frac{1}{2}\left(x^2 + y^2 \right),
\label{pot}
\end{equation}
where
\begin{align}
&r_1 = \sqrt{\left(x + \mu\right)^2 + y^2}, \nonumber\\
&r_2 = \sqrt{\left(x + \mu - 1\right)^2 + y^2},
\label{dist}
\end{align}
are the distances to the respective primaries.

Looking at Eq. (\ref{pot}) we see that the gravitational potential of the first primary body is a classical Newtonian potential of the form $1/r$. The same applies for the second primary only when $p = 1$. When $p < 1$ or $p > 1$ we have the case of a modified Newtonian gravity. We choose $\mu = 1/2$ (which means that both primaries have equal masses) so that the only difference between the two primary bodies to be the power $p$ of the gravitational potential. This choice will help us to unveil the influence of the gravity on the Newton-Raphson basins of attraction since the power $p$ will be the only variable parameter. When $p$ is lower than 1, then the interaction is weaker for short distances and it is stronger for large distances. On the other hand, when $p$ is larger than 1 the interaction is stronger for short distances and it is weaker for larger distances.

The scaled equations of motion describing the motion of the third body in the corotating frame read
\begin{align}
\Omega_x &= \frac{\partial \Omega}{\partial x} = \ddot{x} - 2\dot{y}, \nonumber\\
\Omega_y &= \frac{\partial \Omega}{\partial y} = \ddot{y} + 2\dot{x}.
\label{eqmot}
\end{align}

The dynamical system (\ref{eqmot}) admits the well known Jacobi integral of motion
\begin{equation}
J(x,y,\dot{x},\dot{y}) = 2\Omega(x,y) - \left(\dot{x}^2 + \dot{y}^2 \right) = C,
\label{ham}
\end{equation}
where $\dot{x}$ and $\dot{y}$ are the velocities, while $C$ is the Jacobi constant which is conserved.

\section{Evolution of the equilibrium points}
\label{lgevol}

It was found by Lagrange that five distinct three-body formations exist for two bodies which move in circular orbits around their common center of mass. For an observer in the rotating frame of reference these formations appear to be invariant. Moreover, these special five positions of the test particle for which its location appears to be stationary when viewed from the rotating frame of reference are called Lagrange libration points $L_i$, $i = 1, ..., 5$ \cite{S67}.

\begin{figure*}[!t]
\centering
\resizebox{\hsize}{!}{\includegraphics{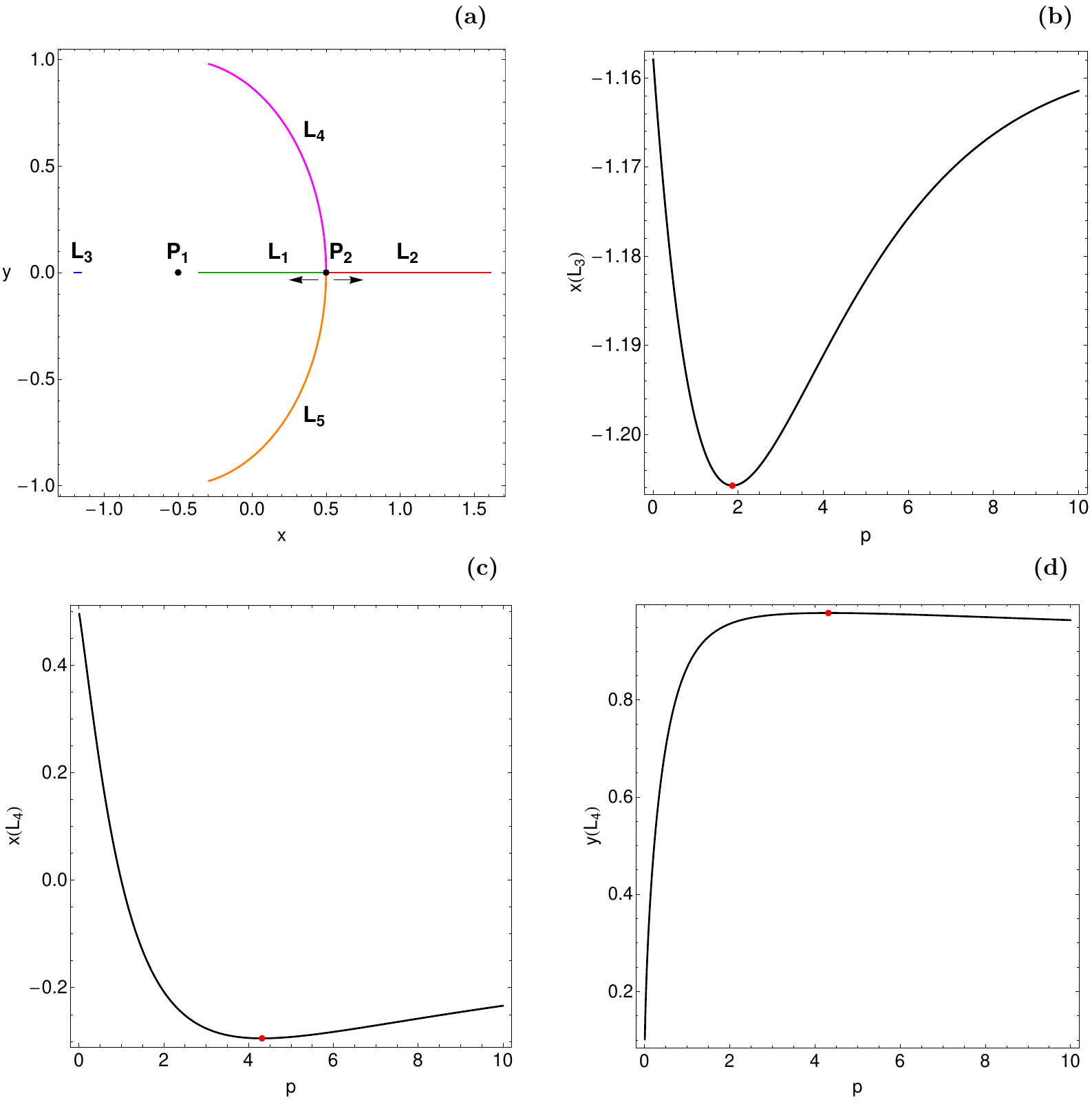}}
\caption{(a-upper left): The space-evolution of the equilibrium points in the planar circular restricted three-body problem with modified Newtonian gravity when $p \in [0.001, 10]$. The arrows indicate the movement direction of the equilibrium points as the power $p$ of the gravitational potential increases. The space-evolution of (b-upper right): the $x$ coordinate of $L_3$, (c-lower left): the $x$ coordinate of $L_4$ and (d-lower right): the $y$ coordinate of $L_4$, as a function of $p$. The red dots pinpoint the turning points of the space-evolution of the coordinates of the Lagrange points.}
\label{lgse}
\end{figure*}

It is well known that in an equilibrium point the following conditions hold
\begin{equation}
\ddot{x} = \ddot{y} = \dot{x} = \dot{y} = 0.
\label{lps1}
\end{equation}
Therefore, the coordinates of the positions of the Lagrange points can be numerically obtained by solving the system of non-linear algebraic differential equations
\begin{equation}
\Omega_x = \Omega_y = 0.
\label{lps}
\end{equation}
Three of the equilibrium points, $L_1$, $L_2$, and $L_3$, (known as collinear points) are located on the $x$-axis, while the other two, $L_4$ and $L_5$, are called triangular points and they are located on the vertices of equilateral triangles. At this point, it should be emphasized that the labeling of the collinear points is not consistent throughout the literature. In this paper, we adopt the most popular case according to which $L_1$ lies between the two primary bodies, $L_2$ is at the right side of $P_2$, while $L_3$ is at the left side of $P_1$ . Therefore we have
\begin{equation}
x(L_3) < - \mu < x(L_1) < 1 - \mu < x(L_2).
\label{pos}
\end{equation}

In this paper, we shall explore how the power $p$ of the gravitational potential of the second primary influences the positions of the libration points. Our results are illustrated in Fig. \ref{lgse}(a-d). In panel (a) we see the space-evolution of the five equilibrium points when $p \in [0.001, 10]$. One may observe that as the power $p$ increases the libration point $L_1$ moves towards the center $P_1$, while the equilibrium point $L_2$ moves away from $P_2$. The space-evolution of the other three equilibrium points on the other hand is not monotonic. In particular, $x(L_3)$ decreases until $p = 1.87$, while for higher values of $p$ it increases. For the coordinates of the Lagrange points $L_4$ and $L_5$ the turning point is at $p = 4.32$. Our numerical calculations suggest that in the limiting case where $p \rightarrow 0$ $L_1$, $L_2$, $L_4$, and $L_5$ tend to collide with the center $P_2$. Here we would like to note that we did not consider cases with negative values of $p$. Our numerical calculations indicate that all five equilibrium points are dynamically unstable when $p \in [0.001, 10]$.

\section{The Newton-Raphson basins of attraction}
\label{bas}

We decided to use the multivariate version of the Newton-Raphson method in order to determine to which of the five equilibrium points each initial point on the configuration $(x,y)$ plane leads to. The Newton-Raphson method is applicable to systems of multivariate functions $f({\bf{x}}) = 0$, through the iterative scheme \begin{equation}
{\bf{x}}_{n+1} = {\bf{x}}_{n} - J^{-1}f({\bf{x}}_{n}),
\label{sch}
\end{equation}
where $J^{-1}$ is the inverse Jacobian matrix of $f({\bf{x_n}})$. In our case the system of equations is
\begin{equation}
\begin{cases}
\Omega_x = 0 \\
\Omega_y = 0
\end{cases},
\label{sys}
\end{equation}
and therefore the Jacobian matrix reads
\begin{equation}
J =
\begin{bmatrix}
\Omega_{xx} & \Omega_{xy} \\
\Omega_{yx} & \Omega_{yy}
\end{bmatrix}.
\label{jac}
\end{equation}
The inverse Jacobian is
\begin{equation}
J^{-1} = \frac{1}{{\rm{det}}(J)}
\begin{bmatrix}
\Omega_{yy} & -\Omega_{xy} \\
-\Omega_{yx} & \Omega_{xx}
\end{bmatrix},
\label{ijac}
\end{equation}
where ${\rm{det}}(J) = \Omega_{yy} \Omega_{xx} - \Omega_{xy}^2$.

Inserting the expression of the inverse Jacobian into the iterative formula (\ref{sch}) we get
\begin{eqnarray}
\begin{bmatrix}
x \\
y
\end{bmatrix}
_{n+1} &=&
\begin{bmatrix}
x \\
y
\end{bmatrix}
_{n} - \frac{1}{\Omega_{yy} \Omega_{xx} - \Omega_{xy}^2}
\begin{bmatrix}
\Omega_{yy} & -\Omega_{xy} \\
-\Omega_{yx} & \Omega_{xx}
\end{bmatrix}
\begin{bmatrix}
\Omega_x \\
\Omega_y
\end{bmatrix}
_{(x_n,y_n)}
\nonumber\\
&=&
\begin{bmatrix}
x \\
y
\end{bmatrix}
_{n} - \frac{1}{\Omega_{yy} \Omega_{xx} - \Omega_{xy}^2}
\begin{bmatrix}
\Omega_{yy}\Omega_x - \Omega_{xy}\Omega_y \\
- \Omega_{yx}\Omega_x + \Omega_{xx}\Omega_y
\end{bmatrix}
_{(x_n,y_n)}.
\label{sch2}
\end{eqnarray}

Decomposing formula (\ref{sch2}) into $x$ and $y$ we obtain the iterative formulae for each coordinate
\begin{eqnarray}
x_{n+1} &=& x_n - \left( \frac{\Omega_x \Omega_{yy} - \Omega_y \Omega_{xy}}{\Omega_{yy} \Omega_{xx} - \Omega^2_{xy}} \right)_{(x_n,y_n)}, \nonumber\\
y_{n+1} &=& y_n + \left( \frac{\Omega_x \Omega_{yx} - \Omega_y \Omega_{xx}}{\Omega_{yy} \Omega_{xx} - \Omega^2_{xy}} \right)_{(x_n,y_n)},
\label{nrm}
\end{eqnarray}
where $x_n$, $y_n$ are the values of the $x$ and $y$ variables at the $n$-th step of the iterative process, while the subscripts of $\Omega$ denote the corresponding partial derivatives of the potential function.

\begin{figure}[!t]
\centering
\includegraphics[width=0.60\hsize]{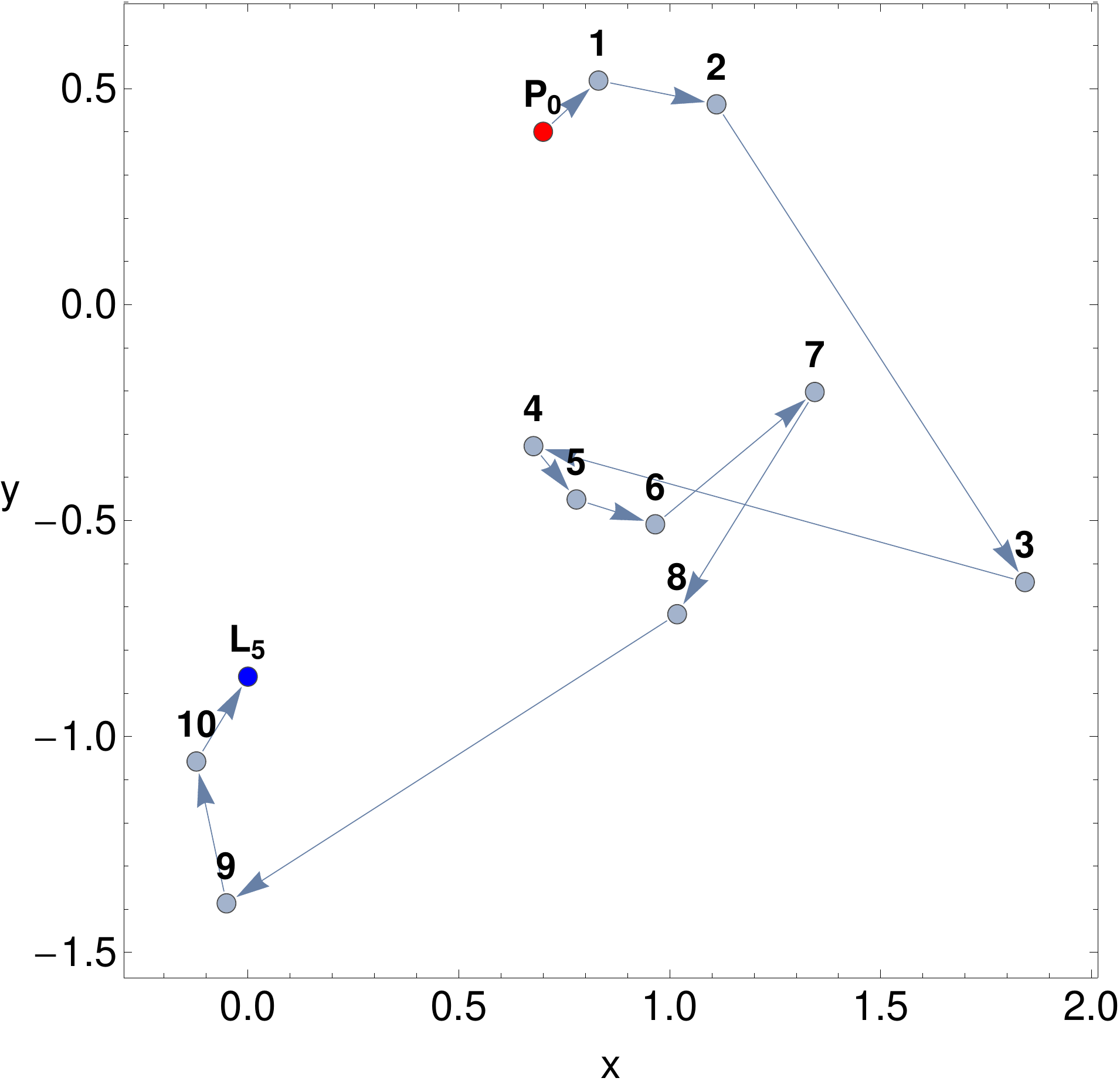}
\caption{A characteristic example of the consecutive steps that are followed by the Newton-Raphson iterator and the corresponding crooked path-line that leads to an equilibrium point $(L_5)$. The red dot indicates the starting point $P_0$, $(x_0, y_0) = (0.7, 0.4)$, while the blue dot indicates the Lagrange point to which the method converged to. For this particular set of initial conditions the Newton-Raphson method converges after 11 iterations to $L_5$ with accuracy of eight decimal digits, while only three more iterations are required for obtaining the desired accuracy of $10^{-15}$.}
\label{crl}
\end{figure}

\begin{figure*}[!t]
\centering
\resizebox{\hsize}{!}{\includegraphics{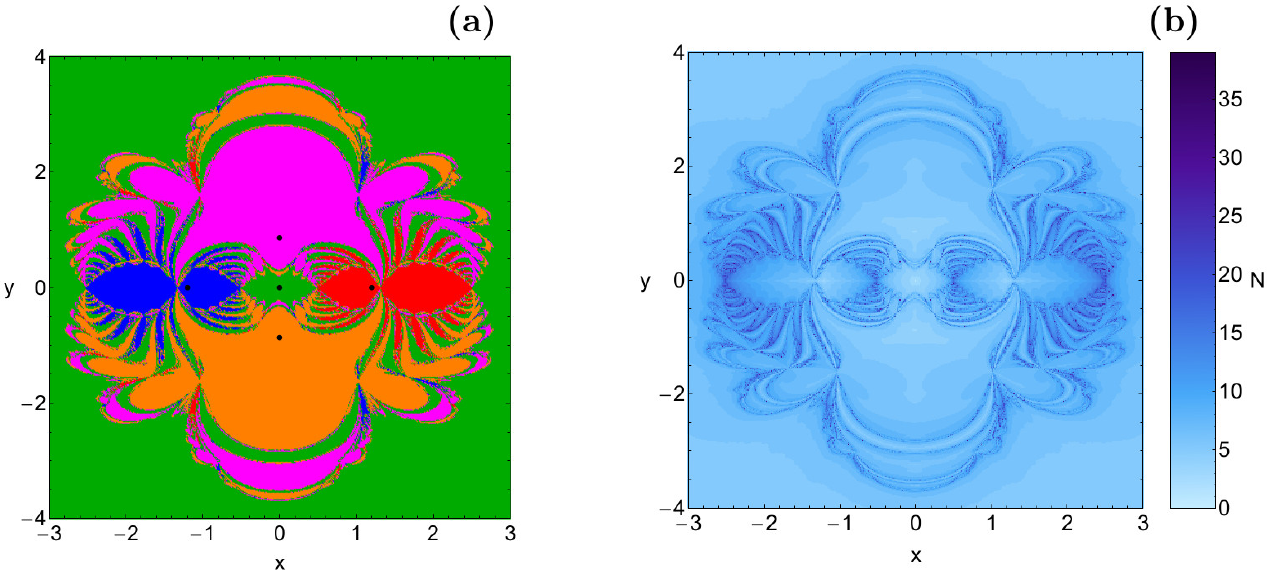}}
\caption{(a-left): The Newton-Raphson basins of attraction on the configuration $(x,y)$ plane for the case of the classical Newtonian gravity $(p = 1)$. The positions of the five equilibrium points are indicated by black dots. The color code denoting the five attractors (Lagrange points) is as follows: $L_1$ (green); $L_2$ (red); $L_3$ (blue); $L_4$ (magenta); $L_5$ (orange); non-converging points (white). (b-right): The distribution of the corresponding number $(N)$ of required iterations for obtaining the Newton-Raphson basins of attraction shown in panel (a).}
\label{sm}
\end{figure*}

\begin{figure}[!t]
\centering
\includegraphics[width=0.60\hsize]{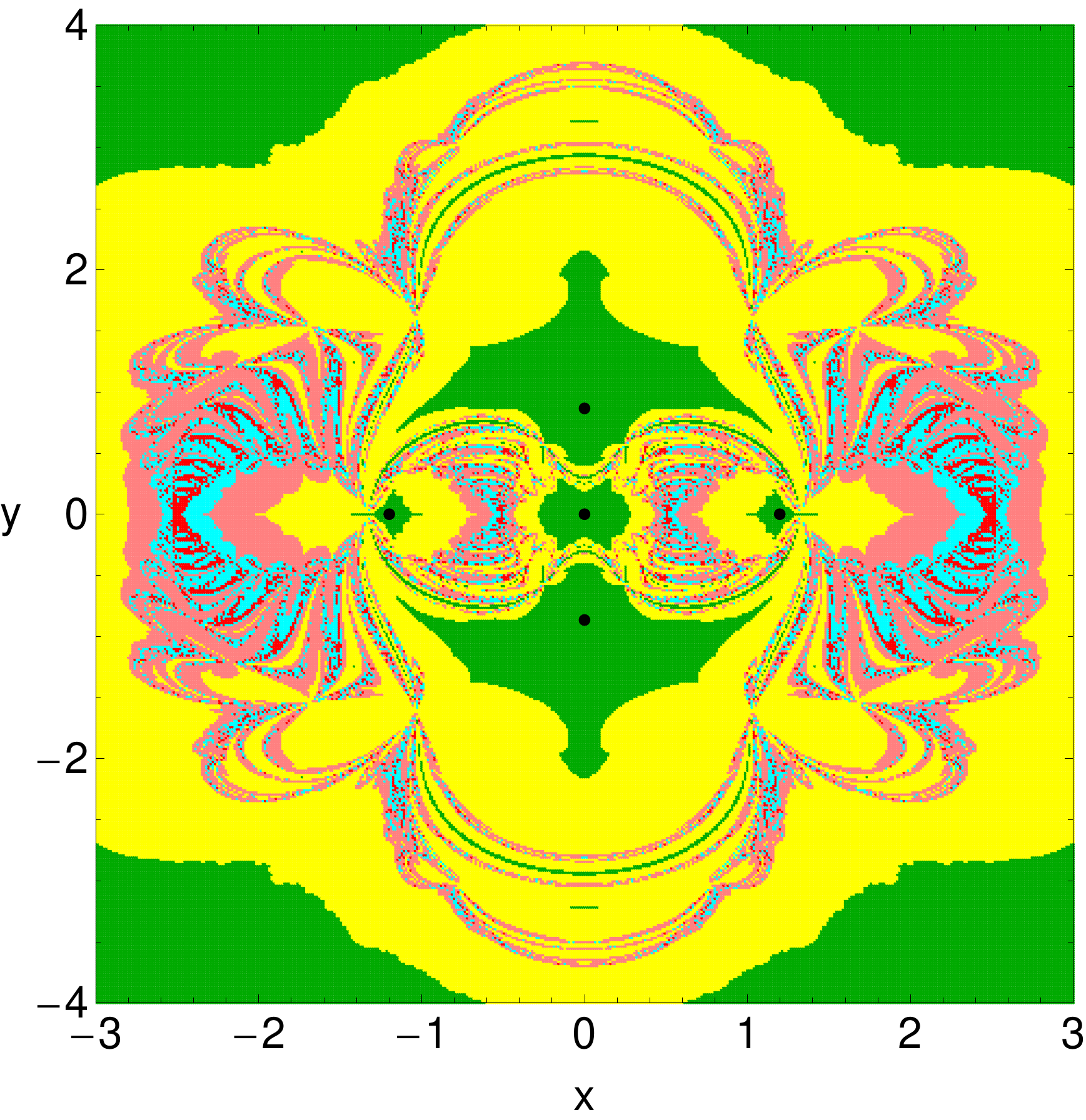}
\caption{Color scale of the Newton-Raphson basins of attraction as a function of the number $N$ of required iterations. The color code is as follows: 0-5 iterations (green); 6-8 iterations (yellow); 9-11 iterations (pink); 12-15 iterations (cyan); $N > 15$ (red). The positions of the five equilibrium points are indicated by black dots.}
\label{iters}
\end{figure}

The Newton-Raphson algorithm is activated when an initial condition $(x_0,y_0)$ on the configuration plane is given, while it stops when the positions of the equilibrium points are reached, with some predefined accuracy. A crooked path line (see Fig. \ref{crl}) is created by the successive approximations-points. If the iterative method converges for the particular initial conditions then this path leads to a desired position, which in our case is one of the five equilibrium points. All the initial conditions that lead to a specific equilibrium point, compose a basin of attraction or an attracting region. Here we would like to clarify that the Newton-Raphson basins of attraction should not be mistaken with the classical basins of attraction in dissipative systems. We observe that the iterative formulae (\ref{nrm}) include both the first and the second derivatives of the effective potential function $\Omega(x,y)$ and therefore we may claim that the obtained numerical results directly reflect some of the basic qualitative characteristics of the dynamical system. The major advantage of knowing the Newton-Raphson basins of attraction in a dynamical system is the fact that we can select the most favorable initial conditions, with respect to required computation time, when searching for an equilibrium point.

For obtaining the basins of convergence we worked as follows: First we defined a dense uniform grid of $1024 \times 1024$ initial conditions regularly distributed on the configuration $(x,y)$ space. The iterative process was terminated when an accuracy of $10^{-15}$ has been reached, while we classified all the $(x,y)$ initial conditions that lead to a particular solution (equilibrium point). At the same time, for each initial point, we recorded the number $(N)$ of iterations required to obtain the aforementioned accuracy. Logically, the required number of iterations for locating an equilibrium point strongly depends on the value of the predefined accuracy. In this study we set the maximum number of iterations $N_{max}$ to be equal to 500.

In Fig. \ref{sm}a we present the Newton-Raphson basins of attraction for the case of the classical Newtonian gravity with $p = 1$. In panel (b) of the same figure the distribution of the corresponding number $(N)$ of iterations required for obtaining the desired accuracy is given using tones of blue. Fig. \ref{iters} shows another type of representation of the distribution of the required iterations. It is seen that around the equilibrium points and also far away from them the iterator converges very quickly (after no more than 5 iterations) to one of the attractors. On the other hand, the highest numbers of required iterations (more than 15) have been identified in the vicinity of the fractal basin boundaries.

All the computations reported in this paper regarding the basins of attraction were performed using a double precision algorithm written in standard \verb!FORTRAN 77! (e.g., \cite{PTVF92}). Furthermore, all graphical illustrations have been created using the version 10.3 of Mathematica$^{\circledR}$ \cite{W03}.

In the following we shall try to determine how the power $p$ of the gravitational potential influences the Newton-Raphson basins of attraction, considering two cases regarding the type of the Newtonian gravity.

\subsection{Weaker Newtonian gravity $(p < 1)$}
\label{ss1}

\begin{figure*}[!t]
\centering
\resizebox{0.8\hsize}{!}{\includegraphics{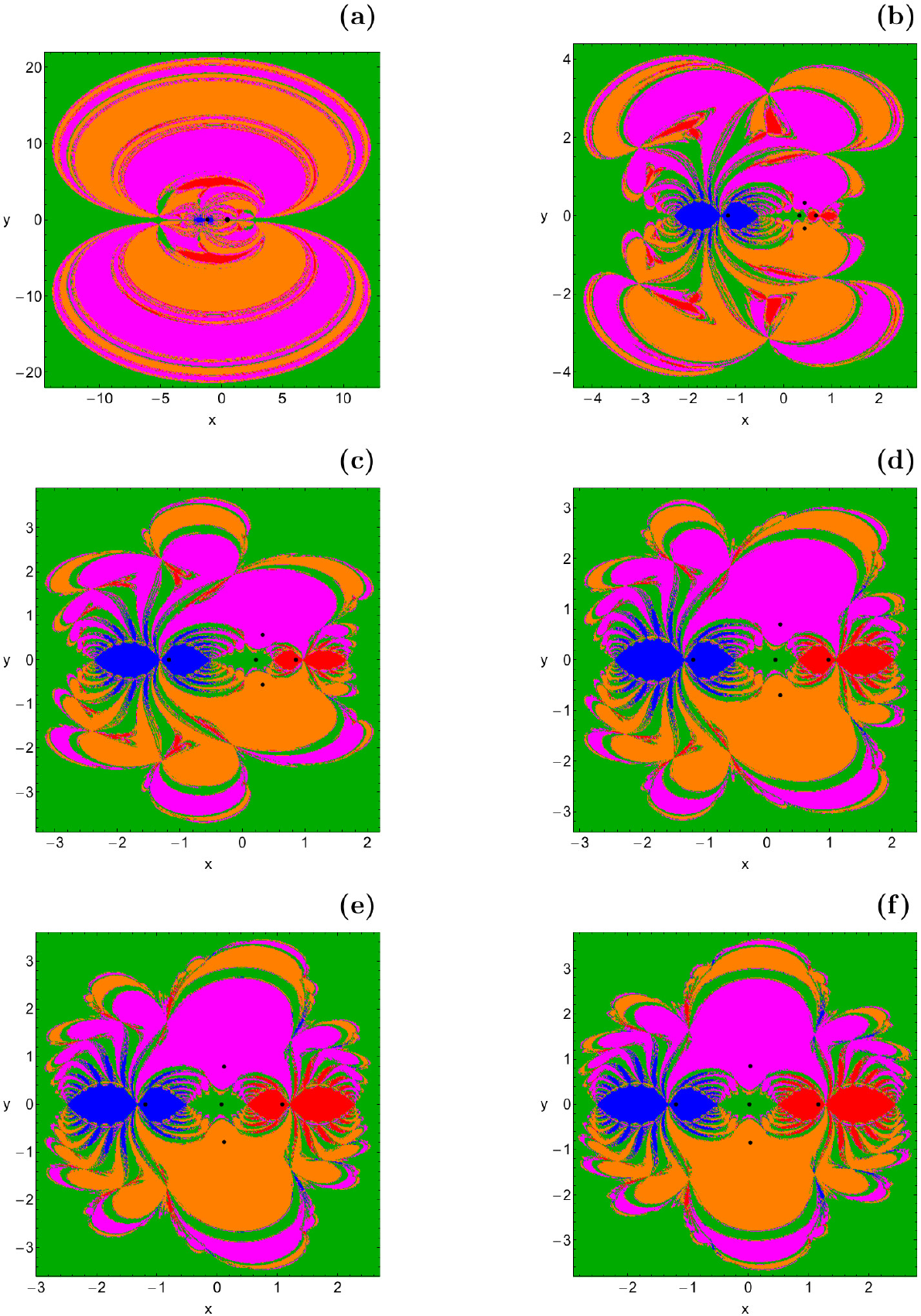}}
\caption{The Newton-Raphson basins of attraction on the configuration $(x,y)$ plane for the case of weaker Newtonian gravity when $p$ varies in the interval $[0.01, 1)$. (a): $p = 0.01$; (b): $p = 0.1$; (c): $p = 0.3$; (d): $p = 0.5$; (e): $p = 0.7$; (f): $p = 0.9$. The positions of the five equilibrium points are indicated by black dots. The color code denoting the five attractors is as in Fig. \ref{sm}a.}
\label{wng}
\end{figure*}

\begin{figure*}[!t]
\centering
\resizebox{0.8\hsize}{!}{\includegraphics{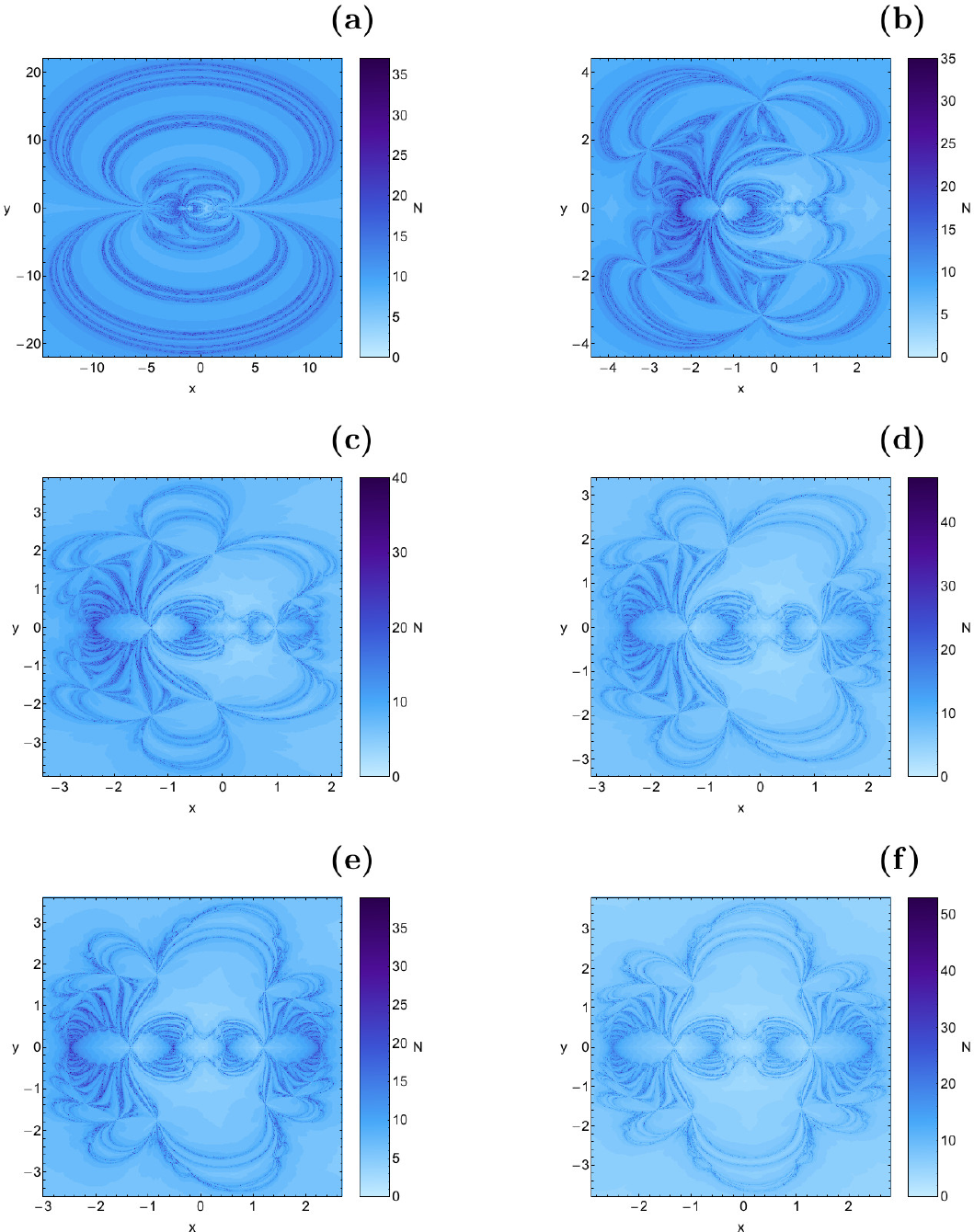}}
\caption{The distribution of the corresponding number $(N)$ of required iterations for obtaining the Newton-Raphson basins of attraction shown in Fig. \ref{wng}(a-f).}
\label{wngn}
\end{figure*}

\begin{figure*}[!t]
\centering
\resizebox{0.8\hsize}{!}{\includegraphics{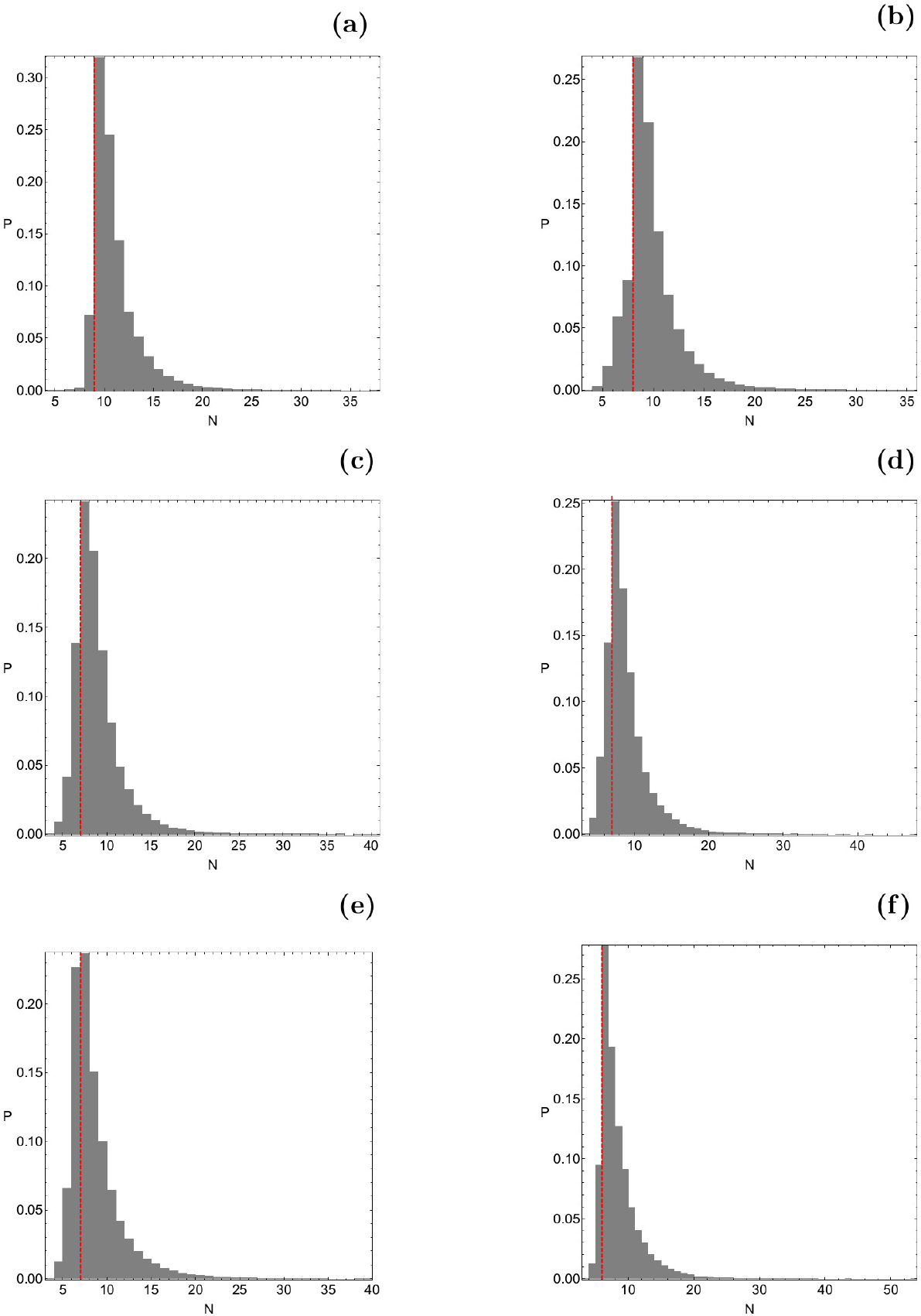}}
\caption{The corresponding probability distribution of required iterations for obtaining the Newton-Raphson basins of attraction shown in Fig. \ref{wng}(a-f). The vertical, dashed, red line indicates, in each case, the most probable number $(N^{*})$ of iterations.}
\label{wngp}
\end{figure*}

\begin{figure*}[!t]
\centering
\resizebox{0.8\hsize}{!}{\includegraphics{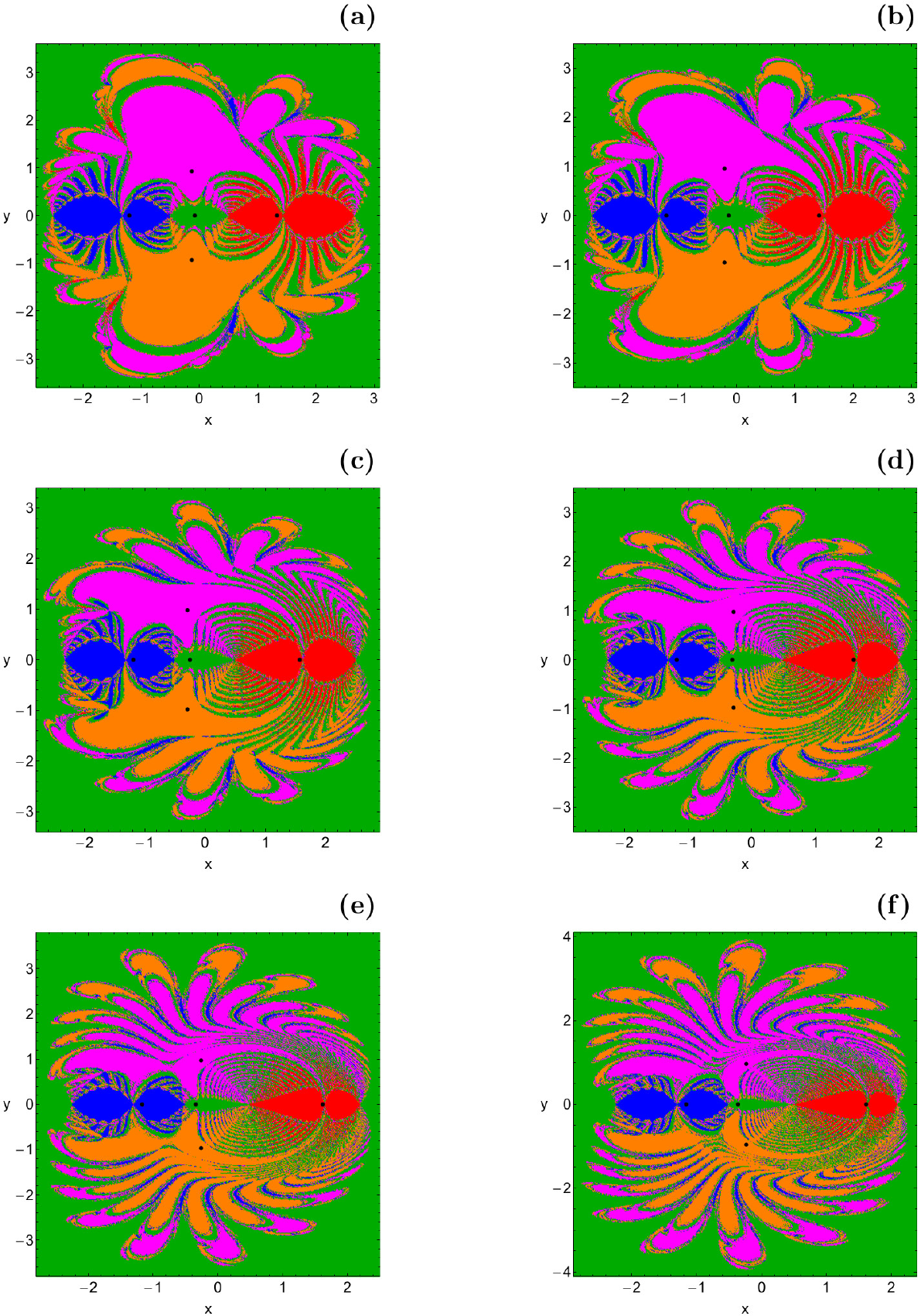}}
\caption{The Newton-Raphson basins of attraction on the configuration $(x,y)$ plane for the case of stronger Newtonian gravity when $p$ varies in the interval $(1, 10]$. (a): $p = 1.5$; (b): $p = 2.0$; (c): $p = 4.0$; (d): $p = 6.0$; (e): $p = 8.0$; (f): $p = 10.0$. The positions of the five equilibrium points are indicated by black dots. The color code denoting the five attractors is as in Fig. \ref{sm}a.}
\label{sng}
\end{figure*}

\begin{figure*}[!t]
\centering
\resizebox{0.8\hsize}{!}{\includegraphics{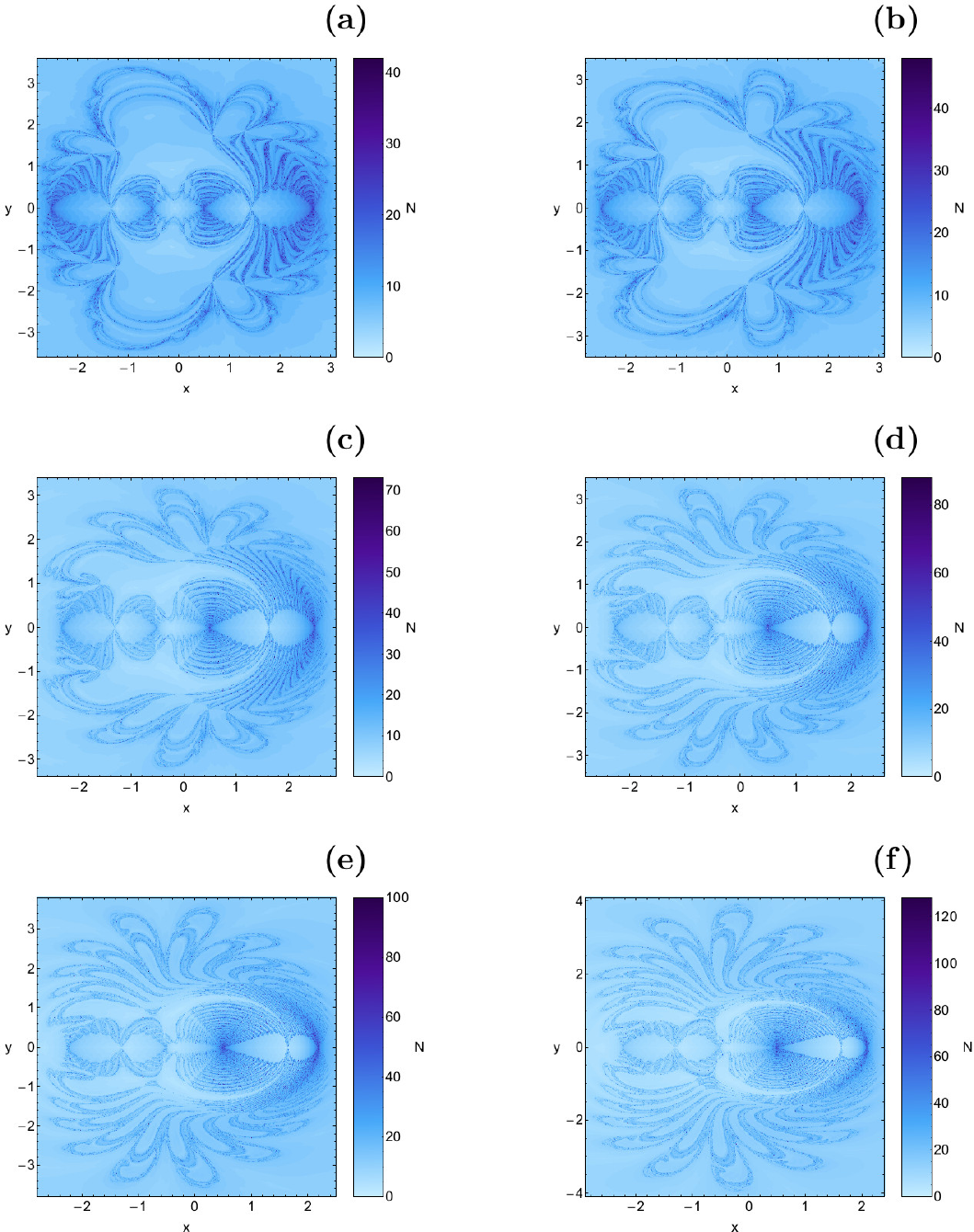}}
\caption{The distribution of the corresponding number $(N)$ of required iterations for obtaining the Newton-Raphson basins of attraction shown in Fig. \ref{sng}(a-f).}
\label{sngn}
\end{figure*}

\begin{figure*}[!t]
\centering
\resizebox{0.8\hsize}{!}{\includegraphics{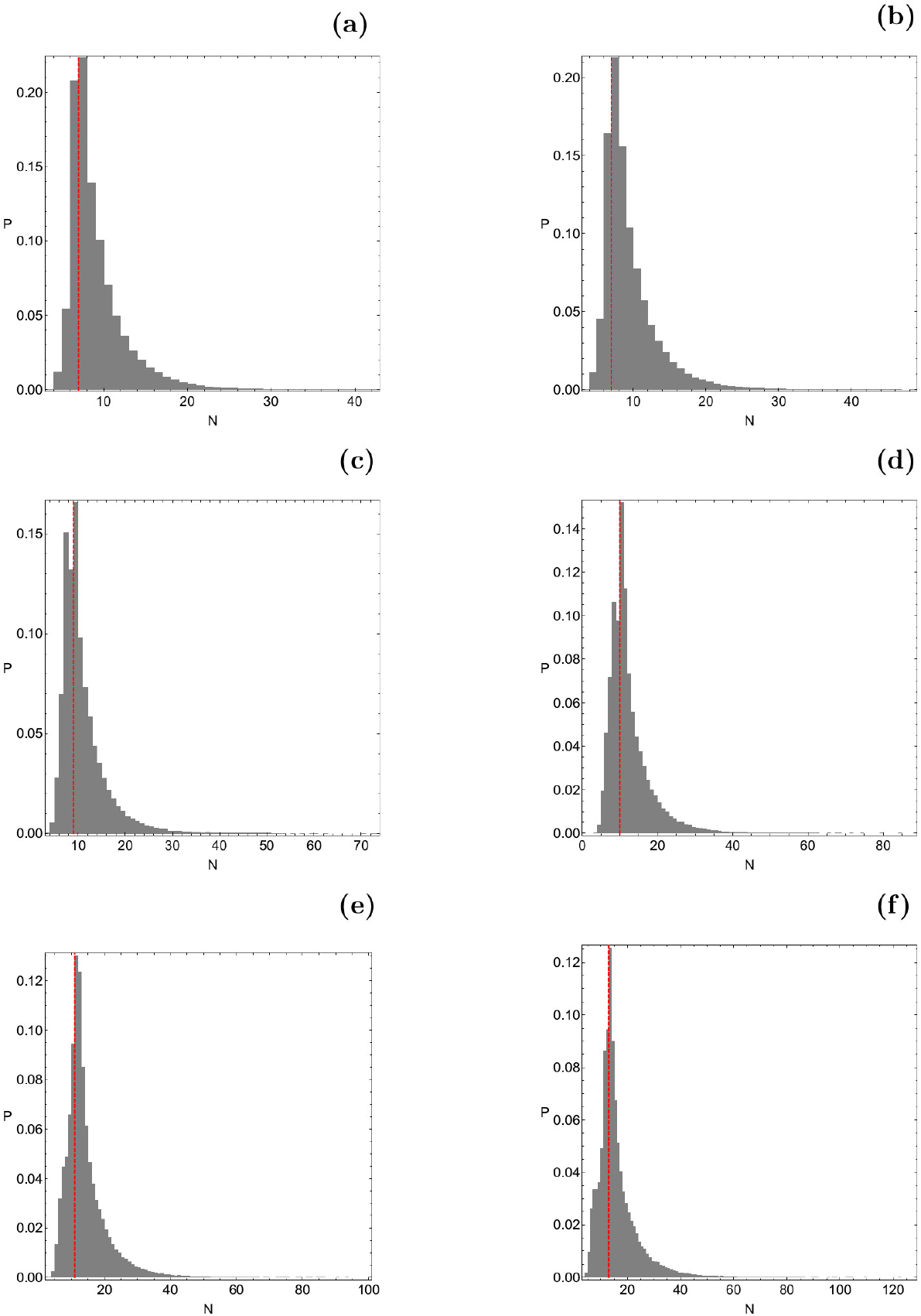}}
\caption{The corresponding probability distribution of required iterations for obtaining the Newton-Raphson basins of attraction shown in Fig. \ref{sng}(a-f). The vertical, dashed, red line indicates, in each case, the most probable number $(N^{*})$ of iterations.}
\label{sngp}
\end{figure*}

Our investigation begins with the case where the Newtonian gravity is weaker and $p$ varies in the interval $[0.01, 1)$. In Fig. \ref{wng}(a-f) we present the Newton-Raphson basins of attraction for six values of the power $p$ of the gravitational potential of the second primary. Looking the color-coded plots we may say that the shape of the basins of attraction corresponding to equilibrium points $L_2$ and $L_3$ have the shape of bugs with many legs and many antennas. Furthermore, the shape of the basins of attraction corresponding to the triangular equilibrium points $L_4$ and $L_5$ looks like multiple butterfly wings. It is interesting to note that the basins of attraction corresponding to the central $L_1$ extend to infinity, while on the other hand the area of all the other basins of convergence is finite. It is evident that a large portion of the configuration $(x,y)$ plane is covered by well-formed basins of attraction. The boundaries between the several basins of convergence however are highly fractal\footnote{When we state that an area is fractal we simply mean that it has a fractal-like geometry without conducting any specific calculations for computing the fractal dimensions.} and they look like a ``chaotic sea". This means that if we choose a starting point $(x_0,y_0)$ of the Newton-Raphson method inside these fractal domains we will observe that our choice is very sensitive. In particular, a slight change in the initial conditions leads to completely different final destination (different attractor) and therefore the beforehand prediction becomes extremely difficult.

In Fig. \ref{wngn}(a-f) we provide the distribution of the corresponding number $(N)$ of iterations required for obtaining the desired accuracy, using tones of blue. In the same vein, in Fig. \ref{wngp}(a-f) the corresponding probability distribution of iterations is shown. The probability $P$ is defined as follows: let us assume that $N_0$ initial conditions $(x_0,y_0)$ converge to one of the five attractors after $N$ iterations. Then $P = N_0/N_t$, where $N_t$ is the total number of initial conditions in every grid. Table \ref{table1} contains the percentages of the Newton-Raphson basins of attraction for the case of weaker Newtonian gravity. It should be noted that the percentage of the basins of attraction corresponding to $L_1$ is not included because these basins extend to infinity and therefore the percentage has no meaning (it depends on the particular size of the rectangular grid).

\begin{table*}[!ht]
\begin{center}
   \caption{The percentages of the Newton-Raphson basins of attraction for the case of weaker Newtonian gravity. Note that the percentages of the basins of convergence corresponding to Lagrange points $L_4$ and $L_5$ are equal.}
   \label{table1}
   \setlength{\tabcolsep}{10pt}
   \begin{tabular}{@{}rccc}
      \hline
      $p$ & $L_2$ (\%) & $L_3$ (\%) & $L_4$ (\%) \\
      \hline
      0.01 & 3.15 & 0.11 & 34.13 \\
      0.10 & 2.89 & 2.12 & 25.08 \\
      0.30 & 2.21 & 3.63 & 20.94 \\
      0.50 & 3.25 & 4.67 & 23.07 \\
      0.70 & 3.86 & 4.61 & 21.43 \\
      0.90 & 4.40 & 4.64 & 20.65 \\
      \hline
   \end{tabular}
\end{center}
\end{table*}

Correlating all the numerical results given in Figs. \ref{wng}, \ref{wngn}, and \ref{wngp}, as well as in Table \ref{table1} one may reasonably deduce that the most important phenomena which take place as the the Newtonian gravity becomes weaker are the following:
\begin{itemize}
  \item The area on the configuration $(x,y)$ plane covered by basins of attraction corresponding to the equilibrium point $L_4$ displays minor fluctuations around 24\%. On the contrary, the area corresponding to the Lagrange point $L_2$ is reduced in the interval $[0.30, 0.90]$. Moreover, the area covered by initial conditions which converge to the equilibrium point $L_3$ exhibits a constant decrease throughout the interval [0.01, 0.50]. It is interesting to note that in the case of weaker Newtonian gravity the shape of the basins of attraction of the equilibrium points $L_4$ and $L_5$ is highly affected by the power $p$ of the gravitational potential.
  \item The average value of required number $(N)$ of iterations for obtaining the desired accuracy increases when $p \rightarrow 0.01$. Consequently, the the most probable number $(N^{*})$ of iterations is increased from 6 when $p = 0.90$ to 9 when $p = 0.01$. In all examined cases, for more than 95\% of the initial conditions on the configuration $(x,y)$ plane the iterative formulae (\ref{nrm}) need no more than 35 iterations for obtaining the desired accuracy.
\end{itemize}

\subsection{Stronger Newtonian gravity $(p > 1)$}
\label{ss2}

We continue with the case where the power $p$ varies in the interval $(1,10]$, which is of course the case of the stronger Newtonian gravity. The Newton-Raphson basins of attraction for six values of the power $p$ are presented in Fig. \ref{sng}(a-f), while Fig. \ref{sngn}(a-f) shows the distribution of the corresponding number $(N)$ of iterations required for obtaining the desired accuracy. The corresponding probability distribution of iterations is given in Fig. \ref{sngp}(a-f). In Table \ref{table2} we provide the percentages of the Newton-Raphson basins of attraction for the case of stronger Newtonian gravity.

\begin{table*}[!ht]
\begin{center}
   \caption{The percentages of the Newton-Raphson basins of attraction for the case of stronger Newtonian gravity. Note that the percentages of the basins of convergence corresponding to Lagrange points $L_4$ and $L_5$ are equal.}
   \label{table2}
   \setlength{\tabcolsep}{10pt}
   \begin{tabular}{@{}rccc}
      \hline
      $p$ & $L_2$ (\%) & $L_3$ (\%) & $L_4$ (\%) \\
      \hline
      1.5 & 5.89 & 5.17 & 20.39 \\
      2.0 & 6.19 & 5.32 & 19.76 \\
      4.0 & 5.50 & 6.23 & 17.48 \\
      6.0 & 4.53 & 6.33 & 17.85 \\
      8.0 & 3.57 & 6.54 & 18.20 \\
     10.0 & 2.89 & 6.69 & 18.68 \\
      \hline
   \end{tabular}
\end{center}
\end{table*}

Taking into consideration all the numerical outcomes presented in Figs. \ref{sng}, \ref{sngn}, and \ref{sngp} as well as in Table \ref{table2} we could argue that the most important phenomena which take place as the Newtonian gravity becomes stronger are the following:
\begin{itemize}
  \item The area on the configuration $(x,y)$ plane corresponding to the equilibrium point $L_4$ (the same applies also for equilibrium point $L_5$) exhibits small fluctuations around 18\%. For $p > 2$ the area corresponding to equilibrium point $L_2$ seems to decrease. On the other hand, the area occupied by basins of attraction corresponding to Lagrange point $L_3$ is constantly increases throughout the interval [1.5, 10.0]. It should be emphasized that in the case of stronger Newtonian gravity the shape of the basins of attraction of the equilibrium points $L_2$, $L_4$ and $L_5$ is strongly affected by the power $p$ of the gravitational potential. In particular, we see that the number of legs and antennas of the basins of attraction corresponding to Lagrange point $L_2$ are increased, as we proceed to higher values of $p$. The same behaviour also applies for the butterfly wings of the basins of attraction of the equilibrium points $L_4$ and $L_5$.
  \item The average value of required number $(N)$ of iterations for obtaining the desired accuracy increases when $p \rightarrow 10$. Consequently, the the most probable number $(N^{*})$ of iterations grows from 7 when $p = 1.5$ to 13 when $p = 10.0$. In all examined cases, for more than 95\% of the initial conditions on the configuration $(x,y)$ plane the iterative formulae (\ref{nrm}) need no more than 50 iterations for obtaining the desired accuracy.
\end{itemize}

\section{Discussion and conclusions}
\label{disc}

The aim of this paper was to numerically obtain the Newton-Raphson basins of attraction in the planar circular restricted three-body problem with modified Newtonian gravity. The basins of convergence of the five equilibrium points of the dynamical system have been determined with the help of the multivariate version of the Newton-Raphson method. These basins describe how each point on the configuration $(x,y)$ plane is attracted by one of the five attractors. Our thorough and systematic numerical investigation revealed how the position of the equilibrium points and the structure of the basins of attraction are influenced by the power $p$ of the gravitational potential describing the gravity around the second primary body. We also found correlations between the basins of attraction and the distribution of the corresponding required number of iterations.

For the numerical calculations of the sets of the initial conditions on the configuration $(x,y)$ plane, we needed about 3 minutes of CPU time on a Quad-Core i7 2.4 GHz PC, depending of course on the required number of iterations. When an initial condition had converged to one of the five attractors with the predefined accuracy the iterative procedure was effectively ended and proceeded to the next available initial condition.

We obtained quantitative information regarding the Newton-Raphson basins of attraction in the restricted three-body problem with modified Newtonian gravity. The main results of our numerical research can be summarized as follows:
\begin{enumerate}
  \item In all examined cases, the configuration $(x,y)$ plane is a complicated mixture of basins of attraction and highly fractal regions. These regions are the exact opposite of the basins of attraction and they are completely intertwined with respect to each other (fractal structure). This sensitivity towards slight changes in the initial conditions in the fractal regions implies that it is impossible to predict the final state (attractor).
  \item The several basins of attraction are very intricately interwoven and they appear either as well-defined broad regions, as thin elongated bands, or even as spiraling structures. The fractal domains are mainly located in the vicinity of the basin boundaries.
  \item The area of the basins of attraction corresponding to collinear equilibrium points $L_2$, $L_3$ as well as to triangular points $L_4$ and $L_5$ is finite. Additional numerical computations reveal that the area of the basins of convergence corresponding to the central equilibrium point $L_1$ extends to infinity.
  \item Our calculations strongly suggest that all initial conditions on the configuration plane converge, sooner or later, to one of the five attractors of the dynamical system. In other words, we did not encounter any initial condition on the configuration plane which did not converge to one of the attractors.
  \item The iterative method was found to converge relatively fast with initial conditions inside the basins of attraction. On the other hand, the highest numbers of required iterations correspond to initial conditions in the fractal domains.
  \item We found that the change on the value of the power $p$ of the potential describing the gravity of the second primary body mostly influences the shape and the geometry of the basins of attraction corresponding to Lagrange points $L_4$ and $L_5$.
  \item The required number of iterations for obtaining the desired accuracy increases when the Newtonian gravity becomes weaker $(p \rightarrow 0.01)$ or stronger $(p \rightarrow 10)$. Our computations indicate that the multivariate Newton-Raphson iterative scheme converges relatively fast in both cases (weaker and stronger Newtonian gravity).
\end{enumerate}

Taking into account the detailed and novel outcomes of our numerical exploration we may suggest that our computational task has been successfully completed. We hope that the present numerical analysis and the corresponding results to be useful in the field of Newton-Raphson basins of attraction in the restricted three-body problem with modified Newtonian gravity. It is in our future plans to expand our investigation in three dimensions thus revealing the basins of attraction inside the three-dimensional (3D) configuration $(x,y,z)$ space. Furthermore, it would be very interesting to use other iterative schemes of higher order than that of the Newton-Raphson and compare the similarities and differences regarding the structure of the basins of convergence.

\section*{Acknowledgments}

\footnotesize

I would like to thank Dr. Paolo Vagnini for all the illuminating and inspiring discussions during this research. My warmest thanks also go to the two anonymous referees for the careful reading of the manuscript and for all the apt suggestions and comments which allowed us to improve both the quality and the clarity of the paper.


\begin{thebibliography}{}

\bibitem{A16} Abu Arqub, O.: The reproducing kernel algorithm for handling differential algebraic systems of ordinary differential equations. Mathematical Methods in the Applied Sciences \textbf{39} 4549-4562 (2016)

\bibitem{AA14} Abu Arqub, O., Abo-Hammour, Z.: Numerical solution of systems of second-order boundary value problems using continuous genetic algorithm. Information Sciences \textbf{279} 396-415 (2014)

\bibitem{AAM16} Abu Arqub, O., Al-Smadi, M., Momani, S., Hayat, T.: Application of reproducing kernel algorithm for solving second-order, two-point fuzzy boundary value problems. Soft Computing, doi:10.1007/s00500-016-2262-3 (2016)

\bibitem{BP11} Baltagiannis, A.N., Papadakis, K.E.: Equilibrium points and their stability in the restricted four-body problem. Int. J. Bifurc. Chaos \textbf{21}, 2179-2193 (2011)

\bibitem{CK07} Croustalloudi, M., Kalvouridis, T.J.: Attracting domains in ring-type $N$-body formations. Planet. Space Sci. \textbf{55}, 53-69 (2007)

\bibitem{CK13} Croustalloudi, M.N., Kalvouridis, T.J.: The Restricted 2+2 body problem: Parametric variation of the equilibrium states of the minor bodies and their attracting regions, ISRN Astronomy and Astrophysics, Article ID 281849 (2013)

\bibitem{D10} Douskos, C.N.: Collinear equilibrium points of Hill's problem with radiation and oblateness and their fractal basins of attraction. Astrophys. Space Sci. \textbf{326}, 263-271 (2010)

\bibitem{EAA15} El-Ajou, A., Abu Arqub, O., Al-Smadi, M.: A general form of the generalized Taylor's formula with some applications. Applied Mathematics and Computation \textbf{256} 851-859 (2015)

\bibitem{FARA14} Fatoorehchi, H., Abolghasemi, H., Rach, R., Assar, M.: An improved algorithm for calculation of the natural gas compressibility factor via the Hall-Yarborough equation of state. The Canadian Journal of Chemical Engineering \textbf{92}, 2211-2217 (2014)

\bibitem{FAR15} Fatoorehchi, H., Abolghasemi, H., Rach, R.: A new parametric algorithm for isothermal flash calculations by the Adomian decomposition of Michaelis-Menten type nonlinearities. Fluid Phase Equilibria \textbf{395}, 44-50 (2015)

\bibitem{GKK09} Gousidou-Koutita, M., Kalvouridis, T.J.: On the efficiency of Newton and Broyden numerical methods in the investigation of the regular polygon problem of $(N + 1)$ bodies. Appl. Math. Comput. \textbf{212}, 100-112 (2009)

\bibitem{JK98} Janicke, L., Kost, A.: Convergence properties of the Newton-Raphson method for nonlinear problems. IEEE Transactions on Magnetics \textbf{34}, 2505-2508 (1998)

\bibitem{K08} Kalvouridis, T.J.: On some new aspects of the photo-gravitational Copenhagen problem. Astrophys. Space Science \textbf{317}, 107-117 (2008)

\bibitem{KGK12} Kalvouridis, T.J., Gousidou-Koutita, M.Ch.: Basins of attraction in the Copenhagen problem where the primaries are magnetic dipoles. Applied Mathematics \textbf{3}, 541-548 (2012)

\bibitem{KK14} Kumari, R., Kushvah, B.S.: Stability regions of equilibrium points in restricted four-body problem with oblateness effects. Astrophys. Space Science \textbf{349}, 693-704 (2014)

\bibitem{PTVF92} Press, H.P., Teukolsky, S.A, Vetterling, W.T., Flannery, B.P.: Numerical Recipes in FORTRAN 77, 2nd Ed., Cambridge Univ. Press, Cambridge, USA (1992)

\bibitem{SSY08} Shih, T-M., Sung, C-H., Yang, B.: A numerical method for solving nonlinear heat transfer equations. Numerical Heat Transfer, Part B: Fundamentals \textbf{54}, 338-353 (2008)

\bibitem{S67} Szebehely, V.: Theory of Orbits, Academic Press, New York (1967)

\bibitem{W03} Wolfram, S.: The Mathematica Book. Fifth Edition, Wolfram Media, Champaign (2003)

\bibitem{Z16} Zotos, E.E.: Fractal basins of attraction in the planar circular restricted three-body problem with oblateness and radiation pressure. Astrophys. Space Sci. \textbf{361}, article id. 181, 17 pp. (2016)

\end{thebibliography}
\end{document}